\documentclass[10pt,journal,letterpaper,twoside]{IEEEtran}

\usepackage{amsmath,graphicx,cite,dsfont,amssymb}
\usepackage{algpseudocode}
\usepackage{amsthm,amsfonts,multirow,bm,array,setspace}
\usepackage{hyperref}
\usepackage{epstopdf}
\usepackage{textcomp}
\usepackage{algpseudocode}

\newcommand{\bX}{\mathbf{X}}

\newcommand{\by}{\mathbf{y}}
\newcommand{\bz}{\mathbf{z}}
\newcommand{\be}{\mathbf{e}}
\newcommand{\bC}{\mathbf{C}}

\newcommand{\bP}{\mathbf{P}}
\newcommand{\bp}{\mathbf{p}}

\title{Flexible Multiple Base Station Association and Activation
for Downlink Heterogeneous Networks}
\author{Kaiming Shen,~\IEEEmembership{Student Member,~IEEE}, Ya-Feng Liu,~\IEEEmembership{Member,~IEEE},\\ David Yiwei Ding, and Wei Yu,~\IEEEmembership{Fellow,~IEEE}%
\thanks{
Manuscript submitted to IEEE Signal Processing Letters. Date of current version August 08, 2017. This work is supported in part by Natural Sciences and
Engineering Research Council (NSERC) of Canada, in part by Huawei Technologies Canada, and in part by the National Natural Science Foundation of China under Grant 11671419. The associate editor coordinating the review of this manuscript and approving it for publication was Prof. Qing Ling.

Kaiming Shen, David Yiwei Ding, and Wei Yu are with The Edward S.~Rogers
Sr.\ Department of Electrical and Computer Engineering, University of Toronto,
Toronto, ON M5S 3G4, Canada (email: kshen@comm.utoronto.ca; davidyding19@gmail.com; weiyu@comm.utoronto.ca).

Y.-F. Liu is with the State Key Laboratory of Scientific and Engineering
Computing, Institute of Computational Mathematics and Scientific/Engineering
Computing, Academy of Mathematics and Systems Science, Chinese Academy
of Sciences, Beijing 100190, China (e-mail: yafliu@lsec.cc.ac.cn).
}
}

\begin{document}


\maketitle

\begin{abstract}
This letter shows that the flexible association of possibly
multiple base stations (BSs) to each user over multiple frequency
bands, along with the joint optimization of BS transmit power that
encourages the turning-off of the BSs at off-peak time, can significantly
improve the performance of a downlink heterogeneous wireless cellular
network. We propose a gradient projection algorithm for optimizing BS
association and an iteratively reweighting scheme together with a novel
proximal gradient method for optimizing power in order to find the optimal
tradeoff between network utility and power consumption. Simulation
results reveal significant performance improvement as compared to
the conventional single-BS association. 
\end{abstract}

\section{Introduction}
\label{sec:intro}

\IEEEPARstart{H}{eterogeneous} network (HetNet), in which the pico base stations (BSs)
are deployed throughout the network to offload traffic from macro BSs,
is a key technique for densifying the BS deployments, thereby improving
the throughput of wireless cellular network. A central question in the
design of HetNet is the joint optimization of cell loading (i.e.,
user-BS association), transmit power, and frequency partitioning
across the pico and macro BSs. In contrast to conventional HetNet
design in which each user is associated with a single BS, this letter
proposes a novel network utility maximization formulation that allows
each user to associate with {\it multiple} BSs over multiple frequency
bands.

The optimization of user-BS association is
inextricably related to power control. Toward this end, this letter
further formulates a BS transmit power spectrum optimization problem
by incorporating a power consumption model that encourages turning off
BSs at off-peak time, and illustrates that the proposed flexible
multiple-BS association and BS activation scheme that takes into account
signal strength, interference management, frequency allocation, and
load balancing can achieve the right balance between network utility
and power consumption in HetNets---although at a cost of larger overhead for coordinated scheduling and power control across the BSs.


This letter solves the proposed network optimization problem
from a group sparse optimization perspective, and uses block coordinate
ascent to iteratively optimize user-BS association and BS transmit
power. We show that the optimization of user-BS association across
the frequencies can be formulated as a convex optimization under fixed
power, and further propose an efficient gradient projection
implementation based on its Lagrangian dual. Moreover, we propose
a \emph{proximal gradient method} for power optimization with a
nonsmooth power penalty. The proposed approach has an faster implementation than
greedy heuristics yet achieves comparable performance.





Wireless cellular networks are traditionally designed with each user
assigned to the BS with the strongest signal, and the BS schedules all
associated users according to some system objective.
It is well known, however, that association based on maximum
signal-to-interference-plus-noise ratio (SINR) may not be optimal, as
it typically under-utilizes pico BSs with lower transmit powers.
The optimal user-BS association problem has been treated from
global optimization \cite{ruoyu,qian} and game theoretic \cite{ha}
perspectives. The most common heuristic used in practice is {\it cell-range
expansion} \cite{lopez, madan}, which can be interpreted as a dual
optimization method that improves upon max-SINR association by taking
cell-loading into account \cite{ye, shen}.
The main observation that underpins the above dual optimization
interpretation is that giving equal time/bandwidth to the multiple
users associated with a single BS is optimal for proportional fairness
\cite{ye}. This observation is true, however, only when each
user is associated with only one BS and all BSs transmit at fixed and
flat power spectral density (PSD) across
the frequencies. The main point of this paper is that
providing the BSs with the flexibility of varying PSDs across the
frequencies and giving the users the flexibility of associating
with possibly multiple BSs over multiple frequency bands can
significantly improve the network utility, albeit at cost of larger
scheduling overhead. Further, significant power saving can be obtained
assuming a realistic power consumption model.




\section{Problem Formulation}

\label{sec:setup}

Consider a downlink heterogeneous cellular
network consisting of $K$ users and $L$ BSs. The total frequency band $W$
is divided into $N$ equal bands. Each user makes a decision at
each band which BS(s) it wishes to associate with. Multiple users
associated with the same BS in the same band are served in a
time-division multiplex fashion. Let $p_\ell^n$ be the
transmit power at BS $\ell$ in band $n$. If user $k$ is served by BS
$\ell$ in band $n$, then its long-term average spectral efficiency is
\begin{equation}
\label{eq:rate}
r^n_{k\ell} = \frac{W}{N}\log \left(1 +
\frac{g_{k\ell}^n p^n_\ell}{\sigma^2 + \sum_{\ell' \neq \ell} g_{k\ell'}^n p_{\ell'}^n}
\right)
\end{equation}
where $\sigma^2$ is the additive noise power, and $g_{k\ell}^n$ is the
long-term channel magnitude from BS $\ell$ to user $k$ in band $n$.

The BS power consumption is modeled as comprised of two parts: the
transmit power $p_{\ell}^n$, and an \textit{on-power} $\psi_\ell$,
which incentivizes BS turning-off during off-peak time:\begin{equation}
\label{eq:powermodel}
Q_\ell(\mathbf p_\ell) = \be_N^T \bp_\ell+\psi_\ell \left\|{\bp_\ell}\right\|_0
\end{equation}
where $\mathbf p_\ell={\left[p^1_\ell,\ldots,p^N_\ell\right]^T},$ $\be_N$ denotes the all-one (column) vector of length $N,$
and $\left\|{\bp_\ell}\right\|_0$ is the zero-norm of vector $\bp_\ell,$ i.e., $\left\|{\bp_\ell}\right\|_0=0$ if $\bp_\ell=\bm{0}$ and $\left\|{\bp_\ell}\right\|_0=1$ otherwise. This BS power consumption model has been adopted in \cite{abbasi,
ping}.  

A main novelty of our problem formulation is that we allow each user
to be associated with potentially different BSs in different frequency
bands.  Let $\bX^n\in\mathbb{R}^{K\times L}$ denote the association
matrix in band $n$ between all users and all BSs. Since multiple users
associated with the same BS and in the same band need to be served with
time-division multiplexing, we use $x_{k\ell}^n$,
the $(k,\ell)$-th entry of $\bX^n$,
to denote the total portion of time user $k$ is served by BS $\ell$ in band $n$.
Then, the transmission rate of user $k$ is
\begin{equation}
R_k=\sum_{\ell=1}^L\sum_{n=1}^N x^n_{k\ell}r^n_{k\ell},~k=1,2,\ldots,K.
\end{equation}
Let $\mathbf P=\left[\mathbf p_1,\mathbf
p_2,\ldots,\mathbf p_L\right]\in\mathbb{R}^{N\times L}$ and let $\bX$
denote the collection of $\left\{\bX^n\right\}_{n=1}^N.$ The objective
is to maximize a network utility, chosen here as a tradeoff between
the BS power consumption and the proportionally fair utility defined
as the sum of log of the user long-term rates:
\begin{subequations}
\label{prob:main}
\begin{eqnarray}
\underset{\mathbf{X},\mathbf P}{\text{max}} &&
\sum_{k=1}^K \log\left(R_k\right) - \lambda \sum_{\ell=1}^L Q_\ell\left(\mathbf p_\ell\right)
  \label{}\\
\text{s.t.}&&
\bm{0}\leq \bp_\ell \leq \bar \bp_\ell,~\ell=1,2,\ldots,L
    \label{prob:main_cons_p}\\
&&\bX^n \be_L \leq  \be_K,~n=1,2,\ldots,N
    \label{prob:main_cons_x1}\\
&&\left(\bX^n\right)^{T} \be_K \leq \be_L,~n=1,2,\ldots,N
    \label{prob:main_cons_x2}\\
&&\bX^n \geq \bm{0},~n=1,2,\ldots,N
    \label{prob:main_cons_x3}
\end{eqnarray}
\end{subequations}
where $\lambda\ge0$ is a constant factor for trade-off purpose, $\bar
\bp_\ell$ is the power budget of BS $\ell$ across all bands.
In problem \eqref{prob:main}, 
$\bX^n \be_L \leq  \be_K$ enforces that the portion of time that
each user is served by all associated BSs should be less than or
equal to one in each band; $\left(\bX^n\right)^{T} \be_K \leq \be_L$
enforces that the portion of time that each BS serves all
associated users should also be less than or equal to one in each band.
It is possible to prove that any $\mathbf{X}^n$ satisfying (\ref{prob:main_cons_x1}) and (\ref{prob:main_cons_x2}) is always realizable via coordinated scheduling across the BSs over multiple time slots \cite{caire}. Note that this formulation differs significantly from prior
works on BS activation and BS association
\cite{ye,shen,caire,sanjabi2014optimal,liao1} in
allowing each user to be served by multiple BSs in different
bands. Further, the BSs may vary PSDs in different bands.



\section{Proposed Algorithms}

\subsection{Iteratively Reweighted Sparse Optimization}

Since the power model \eqref{eq:powermodel} includes a BS on-power that
incentivizes BS turnoff, it is advantageous in the overall problem
formulation \eqref{prob:main} to find a group sparse solution $\bP$, in which
the zero columns correspond to BSs that are turned off.
The paper adopts a nonsmooth mixed $\ell_2/\ell_1$ approach with a
penalty function $\sum_{\ell=1}^L\|\mathbf p_\ell\|_2$ to promote
group sparsity \cite{yuan2006model,shi,liu2016sample}. 
In this case, \eqref{prob:main} can be approximated by the following
weighted mixed $\ell_2/\ell_1$ problem:
\begin{subequations}
\label{eq:foApprox}
\begin{eqnarray}
\underset{\mathbf{X},\mathbf P}{\text{max}} &&
f(\bX,\bP)- \lambda \sum^L_{\ell=1} \psi_\ell w_\ell ||\bp_\ell||_2
  \label{}\\
\text{s.t.}&&
\text{(\ref{prob:main_cons_p})--(\ref{prob:main_cons_x3})}
\end{eqnarray}
\end{subequations}
where
\begin{equation}\label{fxp}
  f(\bX,\bP)=\sum^K_{k=1} \log(R_k) - \lambda \be_L^T\bP\be_N
\end{equation} and $\left\{w_\ell\right\}$ are weights. 
This leads to the following iteratively reweighting algorithm for solving problem \eqref{prob:main}, which consists of solving a sequence of weighted approximation
problems \eqref{eq:foApprox} where the weights used for the next iteration are computed from the value of the current solution.
\begin{center}
\framebox{
\begin{minipage}{8.5cm}
\centerline{\bf Algorithm 1: An Iteratively Reweighting Algorithm}
\centerline{\bf for Solving Problem \eqref{prob:main}}
\vspace{0.05cm} \textbf{Step 1.} Choose a positive sequence $\left\{\tau(t)\right\}.$ Set $t=0$ and $w_\ell(0)=1$ for all $\ell=1,2,\ldots,L.$\\[2.5pt]
\textbf{Step 2.} Solve problem \eqref{eq:foApprox} with $w_\ell=w_\ell(t),~\ell=1,2,\ldots,L$ for its solution $\bP(t)$ and $\bX(t).$\\[2.5pt]
\textbf{Step 3.} Update the weights by
\begin{equation}\label{eq:weightUpdate}
w_\ell(t+1) = \frac{1}{||\bp_\ell(t)||_2 + \tau(t)},~\ell=1,2,\ldots,L,
\end{equation} set $t=t+1,$ and go to \textbf{Step 2.}
\end{minipage}
}
\end{center}

We remark that as
shown in \cite{candes08reweight}, \eqref{eq:weightUpdate} is an
efficient and effective way of updating the weights (for the next
iteration based on the solution of the current one). The idea is to
use smaller weights to penalize BSs with larger $\left\{\|\mathbf p_\ell(t)\|_2\right\}.$ 
In particular, the weights \eqref{eq:weightUpdate} are chosen to be (approximately) inversely proportional to $\left\{\left\|\mathbf p_\ell(t)\right\|_2\right\}$ and $\left\{\tau(t)\right\}$ is introduced to avoid numerical instability. Empirically, it is better to chose $\left\{\tau(t)\right\}$
to be a decreasing sequence than simply setting them to be a small constant value. It has been observed in numerical simulations that Algorithm 1 equipped with \eqref{eq:weightUpdate} converges very fast and often terminates within several iterations.

%
%
%

\subsection{Iterative BS Association and Power Optimization}


It remains to solve the subproblem \eqref{eq:foApprox}. We observe that
since the variables $\bX$ and $\bP$ are separable in the constraints
of \eqref{eq:foApprox}, we can use the block coordinate gradient
ascent (BCGA) algorithm \cite{razaviyayn2013unified,xu2013block,TsengY09} to optimize
BS association and power iteratively.
The proposed BCGA algorithm alternatively updates $\mathbf X$ and
$\mathbf P,$ one at a time with the other being fixed. In the
following, we propose a gradient projection step to update $\bX$
and a proximal gradient ascent step \cite{combettes2011proximal,parikh2014proximal} to update $\bP$ followed by a projection step.
We use $s$ to denote the iteration index in the BCGA algorithm and
$(\bX(s), \bP(s))$ to denote the iterates at the $s$-th iteration.

\subsubsection{Update $\bX$ by Gradient Projection}

At the $(s+1)$-th iteration, the optimization of $\bX$ with $\bP$
fixed as $\bP(s)$ is a convex optimization, because the objective of
\eqref{eq:foApprox} is logarithm of a linear function of $\bX$ and the
constraints are linear. We propose gradient projection as an efficient
method for optimizing $\mathbf X$. The gradient update is:
\begin{equation}
\tilde{\bX}^n(s) = \bX^n(s) + \alpha^n(s) \nabla_{\bX^n}
f\left(\bX(s),\bP(s)\right) 
\end{equation}
where $\alpha^n(s)$ is some appropriately chosen step size.


The projection step is separable among different bands and in
each of band $n=1,2,\ldots,N$ takes the following form
\begin{subequations}
\label{primalx}
\begin{eqnarray}
\underset{\mathbf{X}}{\text{min}} &&
\frac{1}{2} \|\bX^n - \tilde{\bX}^n(s)\|_{\text{F}}^2
  \label{}\\
\text{s.t.}&&
\text{(\ref{prob:main_cons_x1})--(\ref{prob:main_cons_x3})}
\end{eqnarray}
\end{subequations}
where $\|\cdot\|_{\text{F}}$ denotes the Frobenius norm.
The projection step is also a convex optimization problem, but instead of
solving it directly, we show that solving its Lagrangian dual is much
easier.
Let $\by^n$ and $\bz^n$ be the Lagrange multipliers associated with
the linear inequality constraints $\bX^n \be_L \leq  \be_K$ and
$\left(\bX^n\right)^{T} \be_K \leq \be_L,$ respectively. The
Lagrangian dual of problem \eqref{primalx} can be written as
\begin{subequations}
\label{dualx}
\begin{eqnarray}
\underset{\mathbf{y}^n,\mathbf z^n}{\text{min}} &&
\frac{1}{2} \left\|\Theta_s(\by^n,\bz^n)\right\|_{\text{F}}^2+\be_K^T\by^n+\be_L^T\bz^n
  \label{}\\
\text{s.t.}
&& \mathbf y^n\le \bm 0\\
&& \mathbf z^n\le \bm 0
\end{eqnarray}
\end{subequations}
where
$\Theta_s(\by^n,\bz^n)=\max\{\tilde\bX^n(s)+\by^n\be_L^T+\be_K\left(\bz^n\right)^T,\mathbf 0\}$. We relegate the detailed derivation of the above Lagrangian dual to Appendix \ref{appendix:lagrangian}.
The objective function in problem \eqref{dualx} is convex and continuously differentiable over $(\mathbf{y}^n,\mathbf{z}^n)$ and its gradient is
$$\left(\Theta_s(\by^n,\bz^n)\be_L+\be_K,
       \Theta_s(\by^n,\bz^n)^T\be_K+\be_L
   \right)^T.
$$
Since the dual problem (\ref{dualx}) involves only the nonpositive
constraint and the projection of any point onto the nonpositive
orthant is simple to compute, we can easily apply the gradient projection
method to solve the dual problem \eqref{dualx}. After obtaining the
solution $(\by^{n},\bz^{n})$ of the dual problem \eqref{dualx}, we
can recover the solution of the original problem \eqref{primalx} as
\begin{equation}
\label{Xs+1}
\bX^n(s+1)=\Theta_s(\by^{n},\bz^{n}).
\end{equation}

Simulation results show that the above algorithm for solving the
projection problem \eqref{primalx} is much faster than directly
using solver CVX \cite{grant2008cvx}. The above algorithm is an
extension of the DualBB algorithm proposed in \cite[Section
4.3]{jiang2016l_p} from the square matrix case to the non-square
matrix case and from the linear equality constraint case to the linear
inequality constraint case.


\subsubsection{Update $\bP$ by Proximal Gradient}

At the $(s+1)$-th iteration, the optimization of $\bP$ with $\bX$
fixed as $\bX(s+1)$ is nonconvex, but is separable among different BSs.
However, because of the nonsmooth $\ell_2$-norm term in the objective
\eqref{fxp} of problem \eqref{eq:foApprox}, we need to use the proximal
gradient method \cite{combettes2011proximal,parikh2014proximal}. First, compute the gradient update:
\begin{equation}
\label{gradientP}
\tilde{\bp}_{\ell}(s+1) = \bp_{\ell}(s) + \beta_\ell(s+1)
\nabla_{\bp_{\ell}} f\left(\bX(s+1),\bP(s)\right) 
\end{equation}
where $\beta_\ell(s+1)$ is some appropriately chosen step size, e.g., 0.5.
The proximal gradient method \cite{combettes2011proximal,parikh2014proximal}
updates $\mathbf p_\ell$ by solving the following problem for each BS $\ell=1,2,\ldots,L$
\begin{equation}
\label{primalp}
\underset{\bp_{\ell}}{\text{min}}\quad
\frac{1}{2} \left\|\bp_{\ell} -
\tilde{\bp}_{\ell}(s+1)\right\|_2^2 + t_{\ell}(s+1)  \left\|\bp_\ell\right\|_2
\end{equation}
where
\begin{equation}
t_{\ell}(s+1)={\lambda \psi_\ell w_\ell} \beta_\ell(s+1).
\end{equation}
This problem has a closed-form solution \cite[Section 6.5.1]{parikh2014proximal}:
\begin{equation}\label{hatP}\hat \bp_{\ell}(s+1)=\max\left\{1-\frac{t_{\ell}(s+1)}{\left\|\tilde \bp_{\ell}(s+1)\right\|_2}, 0\right\}\tilde\bp_{\ell}(s+1),
\end{equation}
which is sometimes called block soft thresholding. 
We specify
the derivation of the closed-form solution in (15) in Appendix
B for the sake of completeness.
Since $\bp_\ell$ must satisfy the power budget constraint, we further project $\hat \bp_{\ell}(s+1)$ onto the feasible set and obtain the next iterate
\begin{equation}\label{Ps+1}\bp_{\ell}(s+1)=\max\left\{\min\left\{\hat \bp_{\ell}(s+1), \bar \bp_\ell\right\}, \bm{0}\right\}.
\end{equation}

We summarize the proposed BCGA algorithm for solving problem
\eqref{eq:foApprox} as Algorithm 2 below. The initial point can be
chosen as the converged $(\bX,\bP)$ with the previous update of
$w_\ell$.
Global convergence of Algorithm 2 can be established, with
appropriately chosen step sizes $\alpha^n(s)$ and $\beta_\ell(s+1)$
(e.g., by backtracking line search), as in \cite{razaviyayn2013unified,xu2013block,TsengY09}.
\begin{center}
\framebox{
\begin{minipage}{8.5cm}
\centerline{\bf Algorithm 2: A Block Coordinate Gradient Ascent}
\centerline{\bf Algorithm for Solving Problem \eqref{eq:foApprox}}
\vspace{0.05cm} \textbf{Step 1.} Chose an initial point $\left(\bX(0), \bP(0)\right).$ Set $s=0$.\\[2.5pt]
\textbf{Step 2.} Use the gradient projection method to solve problem \eqref{dualx} and use \eqref{Xs+1} to recover the solution $\bX^n(s+1)$ of problem \eqref{primalx} for all $n=1,2,\ldots,N.$\\[2.5pt]
\textbf{Step 3.} Use \eqref{gradientP}, \eqref{hatP} and \eqref{Ps+1} to obtain $\bp_{\ell}(s+1)$ for all $\ell=1,2,\ldots,L.$ Set $s=s+1$ and go to \textbf{Step 2.}
\end{minipage}
}
\end{center}

\subsection{Greedy Algorithm for Solving Problem \eqref{prob:main}}

This subsection outlines a greedy algorithm for solving problem
\eqref{prob:main}, which serves as a benchmark for comparing with
Algorithm 1. The idea is to \emph{sequentially} determine whether or
not to turn off an on-BS with all the other BSs' on-off statuses being fixed. 
More specifically, at each iteration, the greedy algorithm
picks an on-BS $\ell'$ 
and solves problem 
\begin{subequations}
\label{prob:main:smooth}
\begin{eqnarray}
\underset{\mathbf{X},\mathbf P}{\text{max}} &&
f(\mathbf{X},\mathbf P)
  \label{}\\
\text{s.t.}
&& \text{(\ref{prob:main_cons_p})--(\ref{prob:main_cons_x3})}\\
&&\bp_\ell=\bm{0},~\ell\in\cal{L}' \cup \{\ell'\}
\end{eqnarray}
\end{subequations}
where $\cal{L}'$ is the set of off-BSs at the current iteration. The algorithm turns off BS $\ell'$ if the objective value of problem \eqref{prob:main} at the solution of problem \eqref{prob:main:smooth} is larger than that before turning off BS $\ell'$ and otherwise keeps it on. 
Note that problem \eqref{prob:main:smooth} is a smooth version of \eqref{prob:main},
so $\mathbf X$ and $\mathbf P$ can both be optimized by gradient
projection as in Algorithm 2. 
The greedy algorithm for solving
problem \eqref{prob:main} is summarized as Algorithm 3 below.
\begin{center}
\framebox{
\begin{minipage}{8.5cm}
\centerline{\bf Algorithm 3: A Greedy Algorithm}
\centerline{\bf for Solving Problem \eqref{prob:main}}
\vspace{0.05cm} \textbf{Step 1.} Set $\cal{L'}=\emptyset,$ i.e., all BSs are on.\\[2.5pt]
\textbf{Step 2.} Apply Algorithm 2 to optimize $(\mathbf X,\mathbf P)$ for all the on-BSs in \eqref{prob:main:smooth}. \\[2.5pt]
\textbf{Step 3.} Sequentially turn off each on-BS $\ell'$ and set $\cal{L}' = \cal{L}' \cup \{\ell'\}$ if doing so increases the objective value of problem \eqref{prob:main}. Go to \textbf{Step 2.}
\end{minipage}
}
\end{center}

The greedy Algorithm 3 is a simple heuristic, with some
possible drawbacks. First, the greedy algorithm can only turn off (at
most) one BS at each iteration, which makes it inefficient and also
makes it easily to get stuck in a locally optimal solution. In
contrast, the proposed iteratively reweighting Algorithm 1 is capable of
turning off
multiple BSs simultaneously at each iteration; the proposed algorithm
often terminates within only a few iterations. Note that the
performance of Algorithm 3 is sensitive to the sequence of which BS to test in Step 2. In fact, the first few BSs in the testing sequence are more likely to be deactivated than the rest, particularly when the initial utility value is low. A random sequence might lead to quite suboptimal turning-off decisions.

\section{Simulation Results}
\label{sec:res}

\begin{figure}[t!]
\centering
\centerline{\includegraphics[width=\linewidth]{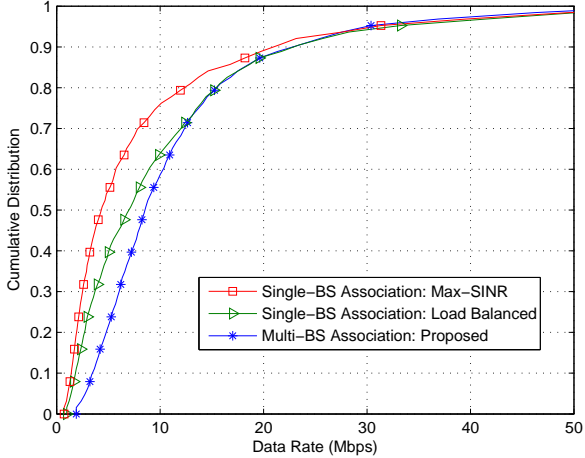}\vspace*{-0.2cm}}
\centering
\caption{Rate CDFs of single-BS vs. multiple-BS association.}
\label{fig:rcdf}\vspace*{-0.35cm}
\end{figure}
We simulate a 7-cell HetNet deployed on a regular hexagon wrapped
around topology, with 1 macro-BS and 3 pico-BSs per cell. A
total of 63 users are uniformly distributed throughout the network.
The downlink distance-dependent pathloss is modeled as
$128.1+37.6\log_{10}d$, where $d$ represents the BS-to-user distance
in km; the shadowing effect is modeled as a zero-mean Gaussian random
variable with 8dB standard deviation. The total
bandwidth is 10MHz, divided into 16 equal bands. The background noise
PSD is -169dBm/Hz; the max transmit PSD of macro-BS is -27dBm/Hz; the
max transmit PSD of pico-BS is -47dBm/Hz.
The on-power of macro-BS is 1450W; the on-power of pico-BS is 21.32W.
Two single-BS association schemes are included as benchmarks:
(i) Max-SINR scheme, which
assigns each user to the BS with the highest SINR then performs power
control under fixed association; (ii) Load balanced scheme \cite{shen},
which uses a proportional fairness objective to associate users to BSs
followed by power control under fixed association. The allocation
variable $\mathbf X$ in these two benchmark schemes is also iteratively
optimized with $\mathbf P$ using Steps 2 and 3 of Algorithm 2.
%
%

Fig.~\ref{fig:rcdf} compares the cumulative distribution function (CDF)
of user rates for multiple-BS vs. single-BS association schemes
without power consumption penalty (i.e., when $\lambda=0$).
It is observed that the proposed multiple-band
multiple-BS association approach significantly outperforms the single-BS
association baselines, particularly in the low-rate regime, e.g.,
the proposed algorithm doubles the 10th percentile rate as compared to
single-BS association. We remark that all the user-BS channels are assumed to be flat-fading across the 16 bands in the simulation, so the reported performance gain is entirely due to multiple-BS association but not frequency diversity.
\begin{figure}[t]
\centering
\includegraphics[width=\linewidth]{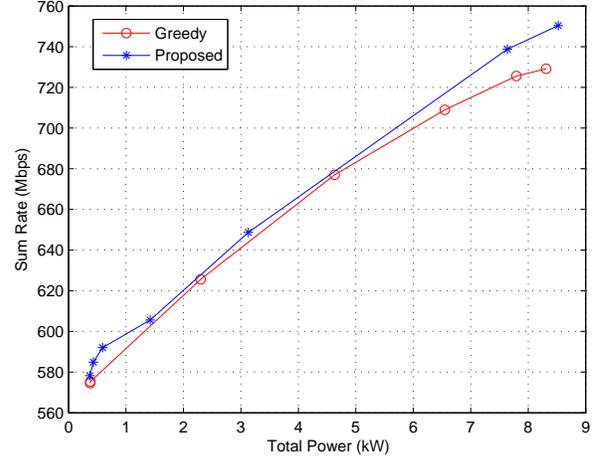}\vspace*{-0.2cm}
\caption{Greedy vs. proposed iteratively reweighting for BS turning-off.}
\label{fig:power_utility}\vspace*{-0.17cm}
\end{figure}

We now evaluate the effectiveness of the proposed algorithms for power
optimization and in turning off BSs for power saving. 
The different points in Fig.~\ref{fig:power_utility} are obtained by
solving (\ref{prob:main}) with different $\lambda$ values. It is observed that the proposed algorithm slightly outperforms the greedy method. 
It should be emphasized that the greedy algorithm is highly
sensitive to the initial starting point and the testing sequence of
BSs (as in Step 3 of Algorithm 3), which are carefully tuned in this
simulation. In particular, we choose to sequentially pick the on-BSs
first across the macro BSs, then across the pico BSs, and terminate
the algorithm until all on-BSs are picked twice.  This testing
sequence is believed to be near-optimal in this particular setting
because the network model has only two tiers of BSs and their
on-power levels differ significantly. In a more complex network, however, it
would be more difficult to find a (near-)optimal
BS testing sequence.
Another drawback of the greedy heuristic is its high complexity,
as it involves testing of candidate BSs one at a time, while our
proposed algorithm is capable of turning off multiple BSs at the same
time through coordinating the proximal updates across the on-BSs.
In fact, the proximal update in our simulation has a fast convergence,
typically within 30 iterations.
As the greedy algorithm
requires a centralized evaluation of the total network utility after
every BS-deactivation attempt, its computational complexity would be prohibitively high when a large number of BSs are
deployed in the network.

\section{Conclusion}
\label{sec:conc}

This letter illustrates the benefit of providing the BSs with the
flexibility of varying transmit PSDs and possibly being turned off, and providing the users with the flexibility of
associating with multiple BSs across multiple frequency bands in a
downlink HetNet. We show that the network utility maximization problem with a power
consumption penalty can be solved efficiently using a combination of
reweighted sparse optimization,
block coordinate ascent, and proximal gradient method.
By jointly optimizing the PSD levels across the bands, the
multiple-BS association, and the user frequency allocation,
the proposed algorithm enables effective balancing between network
throughput and power consumption.


%
\appendices

\section{Lagrangian Dual of Problem (\ref{primalx})}
\label{appendix:lagrangian}

Let $\mathbf y^n\in\mathbb R^{K}$ and $\mathbf{z}^n\in\mathbb R^{L}$ be the Lagrange multipliers associated with the linear inequality constraints (\ref{prob:main_cons_x1}) and (\ref{prob:main_cons_x2}), respectively. The Lagrangian function of problem (\ref{primalx}) can be written as 
\begin{multline}
    {\mathcal L}\left(\mathbf X^n;\mathbf y^n, \mathbf z^n\right) = \frac{1}{2}\left\|\mathbf X^n-\tilde{\mathbf X}^n(s)\right\|^2_\text{F}-\left(\mathbf{X}^n\mathbf{e}_L-\mathbf{e}_K\right)^T \mathbf{y}^n\\
    -\left((\mathbf{X}^n)^T \mathbf{e}_K- \mathbf{e}_L\right)^T \mathbf{z}^n.
\end{multline}
After some manipulations, we can rewrite ${\mathcal L}\left(\mathbf X^n;\mathbf y^n, \mathbf z^n\right)$ as 
 \begin{multline}\label{Lagrangian}
    {\mathcal L}\left(\mathbf X^n;\mathbf y^n, \mathbf z^n\right)=
    \frac{1}{2}\left\|\mathbf X^n\right\|^2_\text{F}-\text{Tr}\left(\mathbf X^n(\tilde{\mathbf X}^n(s))^T\right)-\mathbf e^T_L(\mathbf X^n)^T\mathbf y^n\\
    -\mathbf e^T_K\mathbf X^n \mathbf z^n + \mathbf{e}_K^T \mathbf{y}^n + \mathbf{e}_L^T \mathbf{z}^n+\text{const}(\tilde{\mathbf X}^n(s)),
\end{multline}
 where $\text{Tr}(\cdot)$ is the trace operator and $\text{const}(\tilde{\mathbf X}^n(s))$ denotes a constant term depending only on $\tilde{\mathbf X}^n(s)$.
 The analytic solution of the Lagrangian problem with fixed dual variables, i.e., $\displaystyle \min_{\mathbf X^n\geq \bm{0}}{\mathcal L}\left(\mathbf X^n;\mathbf y^n,\mathbf z^n\right),$ is
\begin{equation}\label{X*}
(\mathbf X^n)^*=\max\left\{\tilde{\mathbf X}^n(s) + \mathbf{y}^n \mathbf{e}^T_L + \mathbf{e}_K( \mathbf{z}^n)^T, \bm{0}\right\}.
\end{equation}
Using the fact $$\frac{1}{2}\left\|\max\left\{\bC, \bm{0}\right\}\right\|^2_\text{F}-\text{Tr}\left(\max\left\{\bC, \bm{0}\right\}\bC^T\right)=-\frac{1}{2}\left\|\max\left\{\bC, \bm{0}\right\}\right\|^2_\text{F}$$ and substituting $\left(\mathbf{X}^n\right)^*$ in \eqref{X*} into  ${\mathcal L}(\mathbf X^n;\mathbf y^n, \mathbf z^n)$ in \eqref{Lagrangian}, we obtain
\begin{multline}
\min_{\mathbf X^n\geq \bm{0}}{\mathcal L}\left(\mathbf X^n;\mathbf y^n,\mathbf z^n\right)\\
=
-\frac{1}{2}\left\|\max\left\{\tilde{\mathbf X}^n(s) + \mathbf{y}^n \mathbf{e}^T_L + \mathbf{e}_K( \mathbf{z}^n)^T,\bm{0}\right\}\right\|^2_\text{F}\\
- \mathbf{e}_K^T \mathbf{y}^n -\mathbf{e}_L^T \mathbf{z}^n +\text{const}(\tilde{\mathbf X}^n).
\end{multline}
Therefore, the following Lagrangian dual of problem (\ref{primalx})
\begin{subequations}
\label{}
\begin{eqnarray}
\underset{\mathbf{y}^n,\,\mathbf z^n}{\text{max}} &&
\min_{\mathbf X^n\geq \bm{0}}{\mathcal L}(\mathbf X^n;\mathbf y^n,\mathbf z^n)
  \label{}\\
\text{s.t.~~}
&& \mathbf y^n\le \bm 0\\
&& \mathbf z^n\le \bm 0
\end{eqnarray}
\end{subequations}
can be equivalently rewritten as problem (\ref{dualx}). 

\section{Closed-Form Solution of Problem \eqref{primalp}}
\label{appendix:proximal}


The optimality condition \cite{analysis} of problem \eqref{primalp} is
\begin{equation}\label{optcondition}
\mathbf p_\ell - \tilde{\mathbf p}_\ell(s+1) + t_{\ell}(s+1) \bm{\xi}=\bm{0},
\end{equation} where $\bm{\xi}\in\partial\|\mathbf p_\ell\|_2$ and $\partial\|\mathbf p_\ell\|_2$ is the subdifferential of the nonsmooth function $\|\mathbf p_\ell\|_2.$
It is simple to compute
  \begin{equation*}
\partial\|\mathbf p_\ell\|_2 = \left\{\begin{array}{cl}
\mathbf p_\ell/\|\mathbf p_\ell\|_2,&\text{if~} \mathbf p_\ell \neq \bm{0};\\[5pt]
\displaystyle \left\{\bm{\xi}\,|\,\|\bm{\xi}\|_2\leq 1 \right\},&\text{if~} \mathbf p_\ell = \bm{0}.
\end{array}
\right.
\end{equation*} Let $\mathbf p^*_\ell$ denote the optimal solution of problem \eqref{primalp}. Now, we consider the following two cases separately.
\begin{itemize}
  \item Case I: $t_{\ell}(s+1)\geq \|\tilde{\mathbf p}_\ell(s+1)\|_2.$ We show that $\mathbf p^*_\ell=\bm{0}$ is the optimal solution of problem \eqref{primalp} in this case. Let $$\bm{\xi}^*=\frac{\tilde{\mathbf p}_\ell(s+1)}{t_{\ell}(s+1)}.$$ Then, $\|\bm{\xi}^*\|_2\leq 1$ and thus $\bm{\xi}^*\in\partial\|\mathbf p^*_\ell\|_2.$ Moreover, it is simple to check that $\mathbf p^*_\ell$ and $\bm{\xi}^*$ satisfy \eqref{optcondition}. This shows that $\mathbf p^*_\ell=\bm{0}$ is the optimal solution of problem \eqref{primalp} when $t_{\ell}(s+1)\geq \|\tilde{\mathbf p}_\ell(s+1)\|_2.$

  \item Case II: $t_{\ell}(s+1) < \|\tilde{\mathbf p}_\ell(s+1)\|_2.$ First, it is simple to argue (by contradiction) that $\bm{0}$ is not the optimal solution of problem \eqref{primalp} in this case. Therefore, the optimality condition of problem \eqref{primalp} can be equivalently rewritten as follows:
      \begin{equation}
\left(\frac{t_{\ell}(s+1)}{\|\mathbf p_\ell\|_2}+1\right)\mathbf p_\ell=\tilde{\mathbf p}_\ell(s+1),
\end{equation} which implies that $\mathbf p^*_\ell$ and $\tilde{\mathbf p}_\ell(s+1)$ are of the same direction, i.e.,
\begin{equation}
\mathbf p^*_\ell = a\tilde{\mathbf p}_\ell(s+1) \text{ for some } a>0.
\end{equation}
With the above form of $\mathbf p^*_\ell$ plugged in, problem (\ref{primalp}) reduces to an optimization problem over the scalar variable $a>0$:
\begin{equation}
\label{}
\underset{a>0}{\text{min}}\quad
\frac{(a-1)^2}{2} \left\|
\tilde{\bp}_{\ell}(s+1)\right\|_2 + at_{\ell}(s+1).
\end{equation}
The closed-form solution of the above problem is
\begin{equation}
a^* = 1-\frac{t_{\ell}(s+1)}{\|\tilde{\bp}_\ell(s+1)\|_2}>0.
\end{equation}
Hence, $\mathbf p^*_\ell=\left(1-\frac{t_{\ell}(s+1)}{\|\tilde{\bp}_\ell(s+1)\|_2}\right)\tilde{\mathbf p}_\ell(s+1)$ is the optimal solution of problem \eqref{primalp} when $t_{\ell}(s+1) < \|\tilde{\mathbf p}_\ell(s+1)\|_2.$
%
%
\end{itemize}

%
%
%
%
%
%
%
From the above cases I and II, we can conclude that (\ref{hatP}) is the solution to problem (\ref{primalp}).
%

\bibliographystyle{IEEEtran}
\bibliography{thesis4}

\end{document}